\documentclass[11pt]{iopart}

\usepackage[usenames]{color}

\newcommand{\bea}{\begin{eqnarray}}
\newcommand{\eea}{\end{eqnarray}}

\begin{document}
%%%%%%%%%%%%%%%%%%%%%%%%%%%%%%%%%%%%%%%%%%%%%%%%%%%%%%%%%%%%%%

%%%%%%%%%%%%%%%%%%%%%%%%%%%%%%%%%%%%%%%%%%%%%%%%%%%%%%%%%%%%%%%
\title{Cosmological hydrodynamics with relativistic pressure and velocity}
\author{Hyerim Noh}
\address{Korea Astronomy and Space Science Institute, Daejeon 305-348, Republic of Korea}
\ead{hr@kasi.re.kr}
\author{Jai-chan Hwang}
\address{Department of Astronomy and Atmospheric Sciences, Kyungpook National University, Daegu 702-701, Republic of Korea}
\ead{jchan@knu.ac.kr}
\author{Chan-Gyung Park}
\address{Division of Science Education and Institute of Fusion Science, Chonbuk National University, Jeonju 561-756, Republic of Korea}
\ead{park.chan.gyung@gmail.com}

%%%%%%%%%%%%%%%%%%%%%%%%%%%%%%%%%%%%%%%%%%%%%%%%%%%%%%%%%%%%%%%
%\date{\today}

%%%%%%%%%%%%%%%%%%%%%%%%%%%%%%%%%%%%%%%%%%%%%%%%%%%%%%%%%%%%%%%
\begin{abstract}

We present hydrodynamic equations with relativistic pressure and velocity in the presence of weak gravity, in a cosmological context. Previously we consistently derived special relativistic hydrodynamic equations with weak gravity in Minkowski background. With the relativistic pressure and velocity one cannot derive the cosmological counterpart by a simple transformation from equations in the Minkowski background. Here we present a proper derivation. We point out the potential importance of relativistic pressure and velocity in gravitational lensing.

\end{abstract}
%%%%%%%%%%%%%%%%%%%%%%%%%%%%%%%%%%%%%%%%%%%%%%%%%%%%%%%%%%%%%%%
\tableofcontents

%%%%%%%%%%%%%%%%%%%%%%%%%%%%%%%%%%%%%%%%%%%%%%%%%%%%%%%%%%%%%%%
\section{Introduction}
                                             \label{sec:Introduction}

In \cite{FNL-2013-HN,FNL-2014-Noh,FNL-2016-multi,FNL-2017-GHNWY} we presented a fully nonlinear and exact perturbation formulation in the background of Robertson-Walker metric. The formulation is powerful in producing higher-order perturbation equations easily, but more interesting aspect is its fully nonlinear and exact nature. We have successfully applied the formulation to Newtonian limit \cite{FNL-2013-Newtonian}, first-order post-Newtonian limit \cite{FNL-2013-PN}, and Newtonian theory with relativistic pressure \cite{FNL-2013-pressure}. In \cite{SRH-2016-HNFPZ,SRH-2016-HN} we applied the formulation to special relativistic hydrodynamics combined with weak gravity in Minkowski background. The resulting equations in these formulations are still fully nonlinear and exact.

By setting the scale factor equal to one and the spatial curvature equal to zero this naturally includes the formulation in the background of Minkowski metric. In this way the formulation in Minkowski background can easily be recovered from the one in Robertson-Walker background, but the opposite is apparently not so simple. In the case of non-relativistic pressure (both isotropic and anisotropic) and velocity (i.e., $p/(\varrho c^2) \ll 1$ and $v^2/c^2 \ll 1$), we can use the equations in Minkowski background to derive the background and perturbation equations in the spatially homogeneous and isotropic but temporally dynamic Robertson-Walker background by using the comoving coordinate ${\bf x}$ (see Sec.\ 9 of \cite{Peebles-1980}) with
\bea
   & & {\bf r} \equiv a(t) {\bf x}, \quad
       \nabla_{\bf r} = {1 \over a} \nabla_{\bf x}, \quad
       {\partial \over \partial t} \bigg|_{\bf r}
       = {\partial \over \partial t} \bigg|_{\bf x}
       - {\dot a \over a} {\bf x} \cdot \nabla_{\bf x}.
\eea
This simple prescription to derive cosmological equations in Newtonian context does not apply in the presence of the relativistic pressure and velocity (cf., \cite{Baqui-2016}). The proper way to derive the background and perturbed equations in the dynamic background is presented in this work.

Our results are the energy and momentum conservation equations and two modified Poisson-type equations in Eqs.\ (\ref{SRH-E-conserv})-(\ref{SRH-Poisson-Psi}). The gravitational lensing potential modified by the presence of relativistic pressure and velocity is presented in Eq.\ (\ref{lensing-potential}). A complete set of fully nonlinear and exact perturbation equations in a flat Robertson-Walker background is summarized in the Appendix. For simplicity we ignore the anisotropic stress.

%%%%%%%%%%%%%%%%%%%%%%%%%%%%%%%%%%%%%%%%%%%%%%%%%%%%%%%%%%%%%%%
\section{Derivation and Results}

%%%%%%%%%%%%%%%%%%%%%%%%%%%%%%%%%%%%%%%%%%%%%%%%%%%%%%%%%%%%%%%
\subsection{Convention}

Our metric convention is
\bea
   & &
       ds^2 = - \left( 1 + {2 \Phi \over c^2} \right) c^2 d t^2
       - 2 \chi_i c dt d x^i
       + a^2 \left( 1 - {2 \Psi \over c^2} \right) \delta_{ij} d x^i d x^j,
   \label{metric}
\eea
where $a (t)$ depends on time only, and $\Phi$, $\Psi$ and $\chi_i$ are general functions of space and time; compared with Eq.\ (\ref{metric-FNL}) we have
\bea
   & & \alpha \equiv {\Phi \over c^2}, \quad
       \varphi \equiv - {\Psi \over c^2},
\eea
where we changed signs of $\Phi$ and $\Psi$ compared with \cite{SRH-2016-HNFPZ,SRH-2016-HN}. The index of $\chi_i$ is raised and lowered by $\delta_{ij}$ as the metric. For vanishing $\Phi$, $\Psi$ and $\chi_i$ we have a {\it flat} Robertson-Walker metric with the scale factor $a(t)$, thus the background medium is homogeneous and isotropic in space; further setting $a \equiv 1$ we recover the Minkowski space-time.

The spatial part of metric is simple because we have {\it ignored} the transverse-tracefree (TT) part (this is a physical assumption, may be acceptable though because TT part corresponds to the gravitational waves to linear order), and have {\it imposed} the spatial gauge condition without losing any generality \cite{Bardeen-1988,2nd-2004-NH,FNL-2013-HN}. Keeping $\chi_i$ is important in the proper analysis; setting the longitudinal part of $\chi_i$ equal to zero (often known as the zero-shear gauge, the longitudinal gauge, or the conformal Newtonian gauge, etc) leads to an {\it inconsistent} result by missing pressure term in the Poisson's equation \cite{SRH-2016-HN}.

We {\it assume} the following two weak gravity conditions and one action-at-a-distance (or small-scale) condition
\bea
   & & {\Phi \over c^2} \ll 1, \quad
       {\Psi \over c^2} \ll 1, \quad
       \gamma^2 {t_\ell^2 \over t_g^2} \ll 1,
   \label{WG-conditions}
\eea
where $t_g$ and $t_\ell$ are gravitational time scale and the light propagating time scale of a characteristic length scale $\ell$, respectively, with
\bea
   & & t_g \sim {1 \over \sqrt{G \varrho}}, \quad
       t_\ell \sim {\ell \over c} \sim {2 \pi a \over kc},
\eea
and $k$ is the comoving wave number with $\Delta = - k^2$; $\gamma$ is the Lorentz factor defined in Eq.\ (\ref{fluid-PT}).
In the dynamic background we have
\bea
   & & {t_\ell \over t_g} \sim {a H \over kc}
       \sim {\ell \over \ell_H}, \quad
       \ell_H \equiv {c \over H}, \quad
       H \equiv {\dot a \over a},
\eea
where $\ell_H$ is often termed as the Hubble horizon in cosmology. Thus the action-at-a-distance condition is stronger than the sub-horizon limit (small-scale limit, $(\ell/\ell_H)^2 \ll 1$) in the case of relativistic speed with $\gamma \gg 1$.

%%%%%%%%%%%%%%%%%%%%%%%%%%%%%%%%%%%%%%%%%%%%%%%%%%%%%%%%%%%%%%%
\subsection{Momentum constraint and the gauge condition}

We decompose $\chi_i \equiv \chi_{,i} + \chi^{(v)}_i$ with $\chi^{(v)i}_{\;\;\;\;\;\;,i} \equiv 0$.
The ADM momentum constraint equation in Eq.\ (\ref{eq3}) becomes
\bea
   & & {2 \over 3} \kappa_{,i}
       + {c \over a^2 {\cal N}}
       \left( {2 \over 3} \Delta \chi_{,i}
       + {1 \over 2} \Delta \chi^{(v)}_i \right)
       = - {8 \pi G \over c^4} a
       \left( \mu + p \right) \gamma^2 v_i.
   \label{eq3-SR}
\eea

We {\it take} the uniform-expansion gauge (UEG, the maximal slicing in Minkowski background) as the temporal gauge (hypersurface or slicing) condition
\bea
   & & \kappa \equiv 0,
\eea
where $\kappa$ is a perturbed part of the trace of extrinsic curvature ($K \equiv K^i_i$) or a perturbed part of the expansion scalar of the normal-frame four vector $\theta^{(n)} \equiv n^c_{\;\; ;c}$ with a minus sign. The UEG together with our spatial gauge condition mentioned below Eq.\ (\ref{metric}) completely removes the gauge degrees of freedom and consequently the remaining variables can be regarded as gauge-invariant variables, to all perturbation orders \cite{Bardeen-1988,2nd-2004-NH,FNL-2013-HN}.

In \cite{SRH-2016-HNFPZ,SRH-2016-HN} we also have considered the zero-shear gauge (often termed as longitudinal or conformal Newtonian gauge) which sets $\chi \equiv 0$ as the temporal gauge condition. In the presence of relativistic pressure this gauge condition leads to a trouble by failing to match with spherically symmetric solution of Tolman-Oppenheimer-Volkoff \cite{Tolman-1939,Oppenheimer-Volkoff-1939}, missing a $3 p$-term in the Poisson's equation, compare Eq.\ (\ref{SRH-Poisson-Phi}) with Eq.\ (\ref{eq4-ZSG-Poisson}). Thus, we concluded that in the presence of relativistic pressure the UEG is the right choice, while both the UEG and the ZSG have proper Newtonian limit. We discuss this issue in more detail in Secs.\ \ref{sec:ZSG}-\ref{sec:UEG-ZSG}.

In our gauge condition, Eq.\ (\ref{eq3-SR}) gives
\bea
   & & \chi = - {12 \pi G \over c^3} a^3 \Delta^{-2} \nabla^i
       \left[ \left( \varrho + {p \over c^2} \right) \gamma^2 v_i \right],
   \label{eq3-SRG-scalar} \\
   & & \chi^{(v)}_i = - {16 \pi G \over c^3} a^3 \Delta^{-1}
       \left\{ \left( \varrho + {p \over c^2} \right) \gamma^2 v_i
       - \Delta^{-1} \nabla_i \nabla^j
       \left[
       \left( \varrho + {p \over c^2} \right) \gamma^2 v_j
       \right] \right\}.
   \label{eq3-SRG-vector}
\eea
Thus
\bea
   & & {1 \over a} \chi_{,i}
       \sim {1 \over a} \chi^{(v)}_i
       \sim \gamma^2 {t_\ell^2 \over t_g^2} {v_i \over c},
\eea
and
\bea
   & & {1 \over a} \chi_{,i}
       \sim {1 \over a} \chi^{(v)}_i
       \ll {v_i \over c}.
   \label{chi-order}
\eea

%%%%%%%%%%%%%%%%%%%%%%%%%%%%%%%%%%%%%%%%%%%%%%%%%%%%%%%%%%%%%%%
\subsection{Conservation equations}

The energy and momentum conservation equations in Eqs.\ (\ref{eq6}) and (\ref{eq7}) give
\bea
   & &
       {d \over dt} \varrho
       + \left( \varrho + {p \over c^2} \right)
       \left[ 3 {\dot a \over a} + {1 \over a} \nabla \cdot {\bf v}
       + {d \over dt} \ln{\gamma} \right]
        = 0,
   \label{eq6-cov-SRG} \\
   & &
       {d \over dt} \left( a \gamma {\bf v} \right)
       + {1 \over \varrho + p/c^2} \left( {1 \over \gamma} \nabla p
       + {1 \over c^2} a \gamma {\bf v} {d \over dt} p \right)
       + \gamma \nabla \Phi
       = 0,
   \label{eq7-cov-SRG}
\eea
where
\bea
   & & {d \over dt} \equiv {\partial \over \partial t}
       + {1 \over a} {\bf v} \cdot \nabla,
\eea
is a convective (Lagrangian) time derivative.
We have $(\ln{\gamma})^{\displaystyle{\cdot}} = \gamma^2 {\bf v} \cdot \dot {\bf v}/c^2$, thus
\bea
   & & {d \over dt} \ln{\gamma}
       = - {\dot a \over a} {v^2 \over c^2}
       - {1 \over \varrho c^2 + p} \left( {d p \over dt}
       - {1 \over \gamma^2} \dot p \right).
   \label{gamma-dot}
\eea

%%%%%%%%%%%%%%%%%%%%%%%%%%%%%%%%%%%%%%%%%%%%%%%%%%%%%%%%%%%%%%%
\subsection{Two Poisson-type equations}

The trace of ADM propagation and the energy constraint equations in Eqs.\ (\ref{eq4}) and (\ref{eq2}), respectively, give
\bea
   & & {\Delta \over a^2} \Phi
       - 3 {\ddot a \over a} + \Lambda c^2
       =
       4 \pi G \left[ \varrho + {3 p \over c^2}
       + 2 \left( \varrho + {p \over c^2} \right)
       \gamma^2 {v^2 \over c^2} \right],
   \label{eq4-SRG} \\
   & & {\Delta \over a^2} \Psi
       + {3 \over 2} \left( {\dot a^2 \over a^2}
       - {\Lambda c^2 \over 3} \right)
       = 4 \pi G \left[ \varrho
       + \left( \varrho + {p \over c^2} \right)
       \gamma^2 {v^2 \over c^2} \right].
   \label{eq2-SRG}
\eea
From these we have
\bea
   & & {\Delta \over a^2} \left( \Phi - \Psi \right)
       - {3 \over 2} \left( 2 {\ddot a \over a}
       + {\dot a^2 \over a^2} - \Lambda c^2 \right)
       = 4 \pi G \left[ {3 p \over c^2}
       + \left( \varrho + {p \over c^2} \right)
       \gamma^2 {v^2 \over c^2} \right].
   \label{eq5-SRG}
\eea
We can show that the tracefree part of ADM propagation in Eq.\ (\ref{eq5}) simply gives Eq.\ (\ref{eq5-SRG}).
Finally, Eq.\ (\ref{eq1}) gives
\bea
   & & c {\Delta \over a^2} \chi
       = {3 \over c^2} \left( \dot \Psi
       + {\dot a \over a} \Phi \right).
   \label{eq1-SR}
\eea
Using Eqs.\ (\ref{eq2-SRG}), (\ref{eq3-SRG-scalar}) and (\ref{eq6-cov-SRG}) we can show that this is naturally valid.
Thus using the complete set of Einstein's equations we have shown the consistency of our relativistic hydrodynamic equations with weak self-gravity in cosmological context: the complete set is Eqs.\ (\ref{eq6-cov-SRG})-(\ref{eq7-cov-SRG}), (\ref{gamma-dot}), (\ref{eq4-SRG}) and (\ref{eq2-SRG}), and $\chi_i$ is determined by Eqs.\ (\ref{eq3-SRG-scalar}) and (\ref{eq3-SRG-vector}).

%%%%%%%%%%%%%%%%%%%%%%%%%%%%%%%%%%%%%%%%%%%%%%%%%%%%%%%%%%%%%%%
\subsection{Equations for background and perturbation}

We decompose density and pressure into background and perturbed parts as
\bea
   & & \varrho \equiv \varrho_b + \delta \varrho, \quad
       p = p_b + \delta p.
\eea
To the background order, Eqs.\ (\ref{eq6-cov-SRG}), (\ref{eq4-SRG})-(\ref{eq5-SRG}) give
\bea
   & & \dot \varrho_b
       + 3 H \left( \varrho_b + {p_b \over c^2}
       \right) = 0,
   \nonumber \\
   & &
       {\ddot a \over a} = - {4 \pi G \over 3}
       \left( \varrho_b
       + 3 {p_b \over c^2} \right)
       + {\Lambda c^2 \over 3},
       \quad
       {\dot a^2 \over a^2}
       = {8 \pi G \over 3} \varrho_b
       + {\Lambda c^2 \over 3}.
   \label{BG-eqs}
\eea
These are well known equations in Friedmann cosmology with flat spatial curvature.

Subtracting the background order, Eqs.\ (\ref{eq6-cov-SRG}), (\ref{eq7-cov-SRG}), (\ref{eq4-SRG}) and (\ref{eq2-SRG}), respectively, give
\bea
   & & {d \over dt} \delta \varrho
       + 3 {\dot a \over a} \left( \delta \varrho + {\delta p \over c^2} \right)
       + \left( \varrho + {p \over c^2} \right)
       \left( {1 \over a} \nabla \cdot {\bf v}
       + {d \over dt} \ln{\gamma} \right)
       = 0,
   \label{SRH-E-conserv} \\
   & &
       {1 \over a \gamma} {d \over dt}
       \left( a \gamma {\bf v} \right)
       = - {1 \over \varrho + p/c^2}
       \left( {1 \over \gamma^2} {1 \over a} \nabla p
       + {1 \over c^2} {\bf v} {d \over dt} p \right)
       - {1 \over a} \nabla \Phi,
   \label{SRH-mom-conserv} \\
   & & {\Delta \over a^2} \Phi
       = 4 \pi G \left[ \delta \varrho + {3 \delta p \over c^2}
       + 2 \left( \varrho + {p \over c^2} \right)
       \gamma^2 {v^2 \over c^2} \right],
   \label{SRH-Poisson-Phi} \\
   & & {\Delta \over a^2} \Psi
       = 4 \pi G \left[ \delta \varrho
       + \left( \varrho + {p \over c^2} \right)
       \gamma^2 {v^2 \over c^2} \right],
   \label{SRH-Poisson-Psi}
\eea
with ${d \over dt} \gamma$ presented in Eq.\ (\ref{gamma-dot}).

By providing the equation of state determining the pressure $p$, Eqs.\ (\ref{SRH-E-conserv})-(\ref{SRH-Poisson-Phi}) together with the background equations are a complete set of equations closed for the variables $\varrho$, $v_i$ and $\Phi$. The other potential $\Psi$ can be determined by Eq.\ (\ref{SRH-Poisson-Psi}) and the remaining metric variable $\chi_i$ can be determined by Eqs.\ (\ref{eq3-SRG-scalar}) and (\ref{eq3-SRG-vector}).

%%%%%%%%%%%%%%%%%%%%%%%%%%%%%%%%%%%%%%%%%%%%%%%%%%%%%%%%%%%%%%%
\subsection{Slow-motion limit}

We consider a slow-motion limit with $v^2/c^2 \ll 1$, thus $\gamma \simeq 1$. From Eqs.\ (\ref{eq6-cov-SRG}), (\ref{eq7-cov-SRG}), (\ref{gamma-dot}), (\ref{SRH-Poisson-Phi}) and (\ref{SRH-Poisson-Psi}) we have \cite{FNL-2013-pressure}
\bea
   & & \dot \varrho
       + 3 {\dot a \over a} \left( \varrho + {p \over c^2} \right)
       + {1 \over a} \nabla \cdot \left( \varrho {\bf v} \right)
       = {1 \over c^2} {1 \over a} \left( {\bf v} \cdot \nabla p
       - p \nabla \cdot {\bf v} \right),
   \label{E-conserv-SM} \\
   & & \dot {\bf v} + {\dot a \over a} {\bf v}
       + {1 \over a} {\bf v} \cdot \nabla {\bf v}
       =
       - {1 \over \varrho + p/c^2}
       \left( {1 \over a} \nabla p
       + {1 \over c^2} \dot p {\bf v} \right)
       - {1 \over a} \nabla \Phi,
   \label{mom-conserv-SM} \\
   & & {\Delta \over a^2} \Phi
       = 4 \pi G \left( \delta \varrho
       + {3 \delta p \over c^2} \right),
   \label{Poisson-Phi-SM} \\
   & & {\Delta \over a^2} \Psi
       = 4 \pi G \delta \varrho.
   \label{Poisson-Psi-SM}
\eea
In the $c \rightarrow \infty$ we recover the Newtonian hydrodynamic equations with gravity \cite{FNL-2013-Newtonian}.

%%%%%%%%%%%%%%%%%%%%%%%%%%%%%%%%%%%%%%%%%%%%%%%%%%%%%%%%%%%%%%%
\subsection{Linear perturbation limit}

To the linear perturbation order, equations in the slow-motion limit are enough. For the longitudinal part, from Eqs.\ (\ref{E-conserv-SM}), (\ref{mom-conserv-SM}) and (\ref{Poisson-Phi-SM}) we can derive
\bea
   & & \ddot \delta
       + \left( 2 + 3 c_s^2 - 6 w \right) H \dot \delta
       + \bigg[
       - 4 \pi G \varrho \left( 1 + w \right)
       \left( 1 + 3 c_s^2 \right)
   \nonumber \\
   & & \qquad
       + 3 \left( c_s^2 - w \right)
       \left( \dot H + 5 H^2 \right)
       + 3 H \left( c_s^2 \right)^{\displaystyle{\cdot}}
       \bigg] \delta
   \nonumber \\
   & & \qquad
       - {\Delta \over a^2} {\delta p \over \varrho}
       = - 3 {1 \over \varrho c^2} \left[ H \dot e
       + \left( 2 \dot H + 5 H^2 \right) e \right],
   \label{delta-eq-UEG}
\eea
where
\bea
   & & w \equiv {p_b \over \mu_b}, \quad
       c_s^2 \equiv {\dot p_b \over \dot \mu_b}, \quad
       \delta p \equiv c_s^2 \delta \mu + e, \quad
       \delta \equiv {\delta \varrho \over \varrho},
\eea
with $\mu_b = \varrho_b c^2$ and $\delta \mu \equiv \delta \varrho c^2$. Equation (\ref{delta-eq-UEG}) can be derived in exactly the same form in the sub-horizon scale (small-scale limit) using the linear perturbation equations in Eqs.\ (132), (134) and (135) of \cite{FNL-2016-multi} in the UEG. Thus we recovered correct linear perturbation limit.

In the zero-pressure limit we recover the well-known equation in the synchronous-comoving gauge \cite{Lifshitz-1946}
\bea
   & & \ddot \delta + 2 {\dot a \over a} \dot \delta
       - 4 \pi G \varrho \delta = 0.
\eea
In the presence of pressure the equation in the comoving gauge \cite{Nariai-1969,Bardeen-1980}, valid in all scales, differs from the one in UEG above. However, as Eq.\ (\ref{delta-eq-UEG}) is derived in the sub-horizon scale, in the presence of relativistic pressure with $\delta p \sim \delta \varrho c^2$, the effectively valid form is
\bea
   & & \ddot \delta
       + \left( 2 + 3 c_s^2 - 6 w \right) {\dot a \over a}
       \dot \delta
       - {\Delta \over a^2} {\delta p \over \varrho}
       = 0,
\eea
which coincides with the equation known in the comoving gauge in the presence of relativistic pressure in the sub-horizon scale \cite{Nariai-1969,Bardeen-1980}.

Now, for the transverse part, from Eq.\ (\ref{mom-conserv-SM}) we have
\bea
   & & \left[ a^4 \left( \varrho + {p \over c^2} \right)
       \nabla \times {\bf v} \right]^{\displaystyle{\cdot}}
       = 0,
\eea
implying the angular momentum conservation $a^4 ( \varrho + {p / c^2} ) \nabla \times {\bf v} = {\bf L} ({\bf x})$ in the absence of anisotropic stress. This equation also follows from the linear perturbation theory, see Eq.\ (140) in \cite{FNL-2016-multi}.

%%%%%%%%%%%%%%%%%%%%%%%%%%%%%%%%%%%%%%%%%%%%%%%%%%%%%%%%%%%%%%%
\subsection{The case in the zero-shear gauge}
                                             \label{sec:ZSG}

We mentioned that in the presence of pressure the ZSG has a problem in reproducing an exact result known in static spherically symmetric system. Here we present the results in the ZSG.

The ZSG sets $\chi \equiv 0$. Equation (\ref{eq3-SR}) gives
\bea
   & & \kappa = - {12 \pi G \over c^2} \Delta^{-1} \nabla \cdot
       \left[ \left( \varrho + {p \over c^2} \right)
       a \gamma^2 {\bf v} \right],
   \label{eq3-ZSG}
\eea
thus
\bea
   & & {a \over c} \Delta^{-1} \nabla \kappa
       \sim {t_\ell^2 \over t_g^2} \gamma^2 {{\bf v} \over c}
       \ll {{\bf v} \over c}.
\eea
Equation (\ref{eq1}) gives
\bea
   & & \kappa = {3 \over c^2} \left( \dot \Psi
       + {\dot a \over a} \Phi \right).
   \label{eq1-ZSG}
\eea

The conservation equations in (\ref{eq6-cov-SRG}) and (\ref{eq7-cov-SRG}) and one of the Poisson-like equation for $\Psi$ in Eq.\ (\ref{eq2-SRG}) remain the same as in the UEG. The {\it only} difference from the UEG appears in the Poisson equation for $\Phi$. Equation (\ref{eq4}) gives
\bea
   & & {\Delta \over a^2} \Phi
       - 3 {\ddot a \over a} + \Lambda c^2
       = - \dot \kappa
       + 4 \pi G \left[ \varrho + {3 p \over c^2}
       + 2 \left( \varrho + {p \over c^2} \right)
       \gamma^2 {v^2 \over c^2} \right].
   \label{eq4-ZSG}
\eea
Compared with Eq.\ (\ref{eq4-SRG}) valid in the UEG, we have $\dot \kappa$ term in the ZSG. Although $\kappa$ term is negligible due to the small-scale limit, it is important to keep $\dot \kappa$ term. Using Eq.\ (\ref{eq3-ZSG}) and the ADM momentum conservation equation in Eq.\ (\ref{eq7-ADM}) we can show that the perturbed part of $3p/c^2$ term is canceled by a term from $\dot \kappa$. Subtracting the background order, we have
\bea
   & & {\Delta \over a^2} \Phi
       = 4 \pi G \left\{ \delta \varrho
       + 2 \left( \varrho + {p \over c^2} \right)
       \gamma^2 {v^2 \over c^2}
       - {3 \over c^2} \Delta^{-1} \nabla_i \nabla_j \left[
       \left( \varrho + {p \over c^2} \right)
       \gamma^2 v^i v^j \right]
       \right\}.
   \label{eq4-ZSG-Poisson}
\eea
Notice that $3 \delta p/c^2$ term is missing compared with Eq.\ (\ref{eq4-SRG}) in the UEG; to the background order the second one in Eq.\ (\ref{BG-eqs}) is correctly reproduced. From Eqs.\ (\ref{eq4-SRG}) and (\ref{eq4-ZSG-Poisson}) we have
\bea
   & & {\Delta \over a^2} \left( \Phi_{\rm UEG}
       - \Phi_{\rm ZSG} \right)
       = {12 \pi G \over c^2} \left\{ \delta p
       + \Delta^{-1} \nabla_i \nabla_j \left[
       \left( \varrho + {p \over c^2} \right)
       \gamma^2 v^i v^j \right]
       \right\}.
   \label{Phi-UEG-ZSG}
\eea

We mentioned that the absence of $3 \delta p$-term in the ZSG fails to reproduce the Tolman-Oppenheimer-Volkoff result in the static spherically symmetric situation. The comparison was made in Sec.\ 2.3 of \cite{SRH-2016-HN}; we note that the $3 p$-term inside an integral in Eq.\ (29) of that paper can be located to the outside the integral as in the exact case of \cite{Oppenheimer-Volkoff-1939} as the difference is higher order in our weak gravity approximation.

%%%%%%%%%%%%%%%%%%%%%%%%%%%%%%%%%%%%%%%%%%%%%%%%%%%%%%%%%%%%%%%
\subsection{Gauge transformation between the UEG and the ZSG}
                                             \label{sec:GT}

We can derive Eq.\ (\ref{Phi-UEG-ZSG}) using the gauge transformation between the UEG and the ZSG. Under the gauge transformation $\widehat x^c = x^c + \xi^c$, to the linear order we have [see Eq.\ (252) in \cite{NH-2004}]
\bea
   & & \widehat \Phi = \Phi - c \dot \xi^t, \quad
       \widehat \Psi = \Psi + H c \xi^t, \quad
       \widehat \kappa = \kappa + \left( 3 \dot H + c^2 {\Delta \over a^2} \right) {1 \over c} \xi^t,
   \nonumber \\
   & & \widehat \chi = \chi - \xi^t, \quad
       \delta \widehat \varrho = \delta \varrho
       - \dot \varrho {1 \over c} \xi^t, \quad
       \delta \widehat p = \delta p - \dot p {1 \over c} \xi^t, \quad
       \widehat v = v - {c \over a} \xi^t.
\eea
We assign the $\widehat x^c$ and $x^c$ coordinates as the UEG and the ZSG, respectively. Thus, using $\widehat \kappa \equiv 0$ and $\chi \equiv 0$ as the respective gauge conditions, we have
\bea
   & & \widehat \chi = - \xi^t, \quad
       \kappa = - \left( 3 \dot H + c^2 {\Delta \over a^2} \right) {1 \over c} \xi^t.
\eea
We can show $\widehat \Psi = \Psi$ by the small-scale limit, $\widehat v = v$ by the weak gravity limit, and $\delta \widehat \varrho = \delta \varrho$ (and $\delta \widehat p = \delta p$) by the weak gravity and small-scale limits (thus difference is doubly suppressed). For $\Phi$ we can show
\bea
   & & \Delta \left( \Phi_{\rm UEG} - \Phi_{\rm ZSG} \right)
       = \Delta \left( \widehat \Phi - \Phi \right)
       = - c \Delta \dot \xi^t
       = \left( a^2 \kappa_{\rm ZSG} \right)^{\displaystyle{\cdot}}
       = c \Delta \dot {\widehat \chi}_{\rm UEG}.
\eea
Using Eq.\ (\ref{eq3-SRG-scalar}) or Eq.\ (\ref{eq3-ZSG}) and using the ADM momentum conservation in Eq.\ (\ref{eq7-ADM}) we can show Eq.\ (\ref{Phi-UEG-ZSG}). We note that although the gauge transformations above are valid only to linear order, it happened that we were able to derive the exact relation in Eq.\ (\ref{Phi-UEG-ZSG}) where the last term is apparently nonlinear.

%%%%%%%%%%%%%%%%%%%%%%%%%%%%%%%%%%%%%%%%%%%%%%%%%%%%%%%%%%%%%%%
\subsection{Comparison of the two gauges in cosmological perturbation}
                                             \label{sec:UEG-ZSG}

Now, let us compare the Poisson equations in the two gauges in the linear cosmological perturbation context. The relevant equations are in Eqs.\ (129)-(133) of \cite{FNL-2016-multi}. We consider a flat background without stress.

In the UEG, the trace of ADM propagation and the ADM energy constraint, respectively, give
\bea
   & & {\Delta \over a^2} \Phi + {3 \over c^2} \dot H \Phi
       = 4 \pi G \left( \delta \varrho + 3 {\delta p \over c^2} \right),
   \label{Poisson-Phi-UEG}
   \\
   & & {\Delta \over a^2} \Psi = 4 \pi G \delta \varrho.
   \label{Poisson-Psi-UEG}
\eea
Thus we have $\Psi \neq \Phi$ even in the small-scale (sub-horizon) limit where Eq.\ (\ref{Poisson-Phi-UEG}) gives the Poisson equation with the pressure term.

In the ZSG, the tracefree ADM propagation gives
\bea
   & & \Psi = \Phi.
\eea
The definition of $\kappa$ and the ADM energy constraint give
\bea
   & & {\Delta \over a^2} \Psi
       - {3 \over c^2} \left( H \dot \Psi + H^2 \Phi \right)
       = 4 \pi G \delta \varrho.
   \label{Poisson-Psi-ZSG}
\eea
In the small-scale limit, these equations give the Poisson equation, now {\it without} the pressure term, see Eq.\ (5.17) in \cite{MFB-1992}.

Although the absence of pressure term in the ZSG contradicts the exact result in the static spherically symmetric system, this does not imply that the ZSG has a serious drawback. This only implies that the variables in the ZSG do not have proper physical meaning (by missing the pressure term), and one has to perform a gauge transformation to variables in other gauge condition, say the UEG in our example. Such a gauge transformation is feasible in the linear perturbation as performed in the previous section. But in the nonlinear situation where our set of equations is valid, such a gauge transformation is not easily available, we somehow succeeded though. In such a case it is necessary to work in the proper gauge from the beginning. On this regards we are proposing the UEG as the proper gauge in the presence of relativistic pressure.

%%%%%%%%%%%%%%%%%%%%%%%%%%%%%%%%%%%%%%%%%%%%%%%%%%%%%%%%%%%%%%%
\section{Discussion}

In this work we have derived weak gravity hydrodynamic equations with relativistic pressure and velocity in cosmological background. These are Eq.\ (\ref{BG-eqs}) for the background and Eqs.\ (\ref{SRH-E-conserv})-(\ref{SRH-Poisson-Phi}) for perturbed part. The remaining metric variables $\Psi$ and $\chi_i$ are determined by Eqs.\ (\ref{SRH-Poisson-Psi}), (\ref{eq3-SRG-scalar}) and (\ref{eq3-SRG-vector}). By setting $a \equiv 1$, $\varrho_b \equiv 0 \equiv p_b$ and $\Lambda = 0$, we recover the special relativistic hydrodynamic equations with weak gravity in Minkowski background \cite{SRH-2016-HN}.

In the weak gravity limit, in Minkowski background, the null geodesic equation gives [see Eq.\ (148) in \cite{HNP-PN-2008}]
\bea
   & & {d^2 {\bf x} \over d t^2}
       = - \nabla \left( \Phi + \Psi \right).
\eea
From Eqs.\ (\ref{SRH-Poisson-Phi}) and (\ref{SRH-Poisson-Psi}) we have
\bea
   & & \Delta \left( \Phi + \Psi \right)
       = 4 \pi G \left[ 2 \varrho + {3 p \over c^2}
       + 3 \left( \varrho + {p \over c^2} \right)
       \gamma^2 {v^2 \over c^2}
       + 3 \Pi_{ij} {v^i v^j \over c^2} \right],
   \label{lensing-potential}
\eea
where we have added a contribution from anisotropic pressure presented in Eqs.\ (5) and (10) of \cite{SRH-2016-HN}; we have $\widetilde \pi_{ij} \equiv \Pi_{ij}$ with indices of $\Pi_{ij}$ raised and lowered by $\delta_{ij}$ as the metric; $\varrho$ includes internal energy density as well as the rest mass density. Thus, the relativistic pressure (both isotropic and anisotropic) as well as velocity may affect the gravitational lensing which could be nonnegligible in the micro-lensing event passing near compact object with significant special relativistic astrophysical processes. Exotic form of pressure as often introduced in the dark energy study could also have a role on weak-lensing and macro-lensing events. In the conventional gravitational lensing only the rest mass density has the role. The research in these fields is open to future investigation.

%%%%%%%%%%%%%%%%%%%%%%%%%%%%%%%%%%%%%%%%%%%%%%%%%%%%%%%%%%%%%%%
%
%  Acknowledgments
%
%%%%%%%%%%%%%%%%%%%%%%%%%%%%%%%%%%%%%%%%%%%%%%%%%%%%%%%%%%%%%%%
\section*{Acknowledgments}
J.H.\ was supported by Basic Science Research Program through the National Research Foundation (NRF) of Korea funded by the Ministry of Science, ICT and future Planning (No.\ 2016R1A2B4007964 and No.\ 2018R1A6A1A06024970). H.N.\ was supported by National Research Foundation of Korea funded by the Korean Government (No.\ 2018R1A2B6002466). C.-G.P. was supported by the Basic Science Research Program through the National Research Foundation of Korea (NRF) funded by the Ministry of Education (No. 2017R1D1A1B03028384).

%%%%%%%%%%%%%%%%%%%%%%%%%%%%%%%%%%%%%%%%%%%%%%%%%%%%%%%%%%%%%%%
\section*{Appendix: Fully nonlinear and exact perturbation equations}

Our metric convention of the fully nonlinear and exact perturbation theory in a flat Friedmann background is \cite{Bardeen-1988,FNL-2013-HN}
\bea
   & & ds^2 = - a^2 \left( 1 + 2 \alpha \right) (d x^0)^2
       - 2 a \chi_{i} d x^0 d x^i
       + a^2 \left( 1 + 2 \varphi \right) \delta_{ij} d x^i d x^j,
   \label{metric-FNL}
\eea
where the spatial index of $\chi_i$ is raised and lowered by $\delta_{ij}$ as the metric. Here we {\it assume} $a$ to be a function of time only, and $\alpha$, $\varphi$ and $\chi_i$ are functions of space and time with arbitrary amplitudes. The spatial part of the metric is simple because we already have taken the spatial gauge condition {\it without} losing any generality, and have {\it ignored} the transverse-tracefree tensor-type perturbation; extension to most general situation without these assumptions can be found in \cite{FNL-2017-GHNWY}. We have not imposed the temporal gauge (slicing) condition, and under our spatial gauge condition together with suitable slicing conditions our perturbation variables are spatially and temporally gauge invariant to fully nonlinear order \cite{Bardeen-1988,2nd-2004-NH,FNL-2013-HN}.

The energy-momentum tensor of a fluid is
\bea
   & & \widetilde T_{ab}
       = \widetilde \mu \widetilde u_a \widetilde u_b
       + \widetilde p \widetilde h_{ab} + \widetilde \pi_{ab},
   \label{Tab}
\eea
where $\widetilde u_a$ is the normalized four-vector with $\widetilde u^a \widetilde u_a \equiv -1$ and $\widetilde h_{ab} \equiv \widetilde g_{ab} + \widetilde u_a \widetilde u_b$ is the projection tensor. Tildes indicate covariant quantities; $\widetilde \mu$, $\widetilde p$ and $\widetilde \pi_{ab}$ are the covariant energy density, pressure and anisotropic stress respectively, with $\widetilde \pi_{ab} \widetilde u^a \equiv 0 \equiv \widetilde \pi^c_c$. We have taken the energy-frame condition without losing any generality thus $\widetilde u_a$ is the fluid four-vector \cite{Ellis-1971,Ellis-1973}. For simplicity we ignore the anisotropic stress in this work.

In the perturbation theory we may introduce \cite{FNL-2013-HN}
\bea
   \fl \widetilde \mu \equiv \mu, \quad
       \widetilde p \equiv p, \quad
       \widetilde u_i \equiv a \gamma {v_i \over c}, \quad
       \gamma \equiv {1 \over \sqrt{ 1
       - {1 \over 1 + 2 \varphi} {v^2 \over c^2}}}, \quad
       {\cal N} \equiv \sqrt{ 1 + 2 \alpha
       + {\chi^k \chi_k \over a^2 ( 1 + 2 \varphi )}},
   \label{fluid-PT}
\eea
where spatial index of $v_i$ is raised and lowered by $\delta_{ij}$ as the metric; the perturbed fluid quantities $\mu$, $p$ and $v_i$ are functions of space and time with arbitrary amplitudes; $\gamma$ is the Lorentz factor with $v^2 \equiv v^k v_k$; $N \equiv a {\cal N}$ is the lapse function. The complete set of fully nonlinear and exact perturbation equations is presented below \cite{FNL-2013-HN,FNL-2014-Noh,FNL-2016-multi}.

\noindent
Definition of $\kappa$: \bea
   \fl \kappa
       \equiv
       3 {\dot a \over a} \left( 1 - {1 \over {\cal N}} \right)
       - {1 \over {\cal N} (1 + 2 \varphi)}
       \left[ 3 \dot \varphi
       + {c \over a^2} \left( \chi^k_{\;\;,k}
       + {\chi^{k} \varphi_{,k} \over 1 + 2 \varphi} \right)
       \right].
   \label{eq1}
\eea
ADM energy constraint:
\bea
   \fl - {3 \over 2} \left( {\dot a^2 \over a^2}
       - {8 \pi G \over 3} \varrho
       - {\Lambda c^2 \over 3} \right)
       + {\dot a \over a} \kappa
       + {c^2 \Delta \varphi \over a^2 (1 + 2 \varphi)^2}
   \nonumber \\
   \fl \qquad
       = {1 \over 6} \kappa^2
       - 4 \pi G \left( \varrho + {p \over c^2} \right)
       \left( \gamma^2 - 1 \right)
       + {3 \over 2} {c^2 \varphi^{,i} \varphi_{,i} \over a^2 (1 + 2 \varphi)^3}
       - {c^2 \over 4} \overline{K}^i_j \overline{K}^j_i.
   \label{eq2}
\eea
ADM momentum constraint:
\bea
   \fl {2 \over 3} \kappa_{,i}
       + {c \over 2 a^2 {\cal N} ( 1 + 2 \varphi )}
       \left( \Delta \chi_i
       + {1 \over 3} \chi^k_{\;\;,ik} \right)
       + 8 \pi G \left( \varrho + {p \over c^2} \right)
       a \gamma^2 {v_{i} \over c^2}
   \nonumber \\
   \fl \qquad
       =
       {c \over a^2 {\cal N} ( 1 + 2 \varphi)}
       \Bigg\{
       \left( {{\cal N}_{,j} \over {\cal N}}
       - {\varphi_{,j} \over 1 + 2 \varphi} \right)
       \left[ {1 \over 2} \left( \chi^{j}_{\;\;,i} + \chi_i^{\;,j} \right)
       - {1 \over 3} \delta^j_i \chi^k_{\;\;,k} \right]
   \nonumber \\
   \fl \qquad
       - {\varphi^{,j} \over (1 + 2 \varphi)^2}
       \left( \chi_{i} \varphi_{,j}
       + {1 \over 3} \chi_{j} \varphi_{,i} \right)
       + {{\cal N} \over 1 + 2 \varphi} \nabla_j
       \left[ {1 \over {\cal N}} \left(
       \chi^{j} \varphi_{,i}
       + \chi_{i} \varphi^{,j}
       - {2 \over 3} \delta^j_i \chi^{k} \varphi_{,k} \right) \right]
       \Bigg\}.
   \label{eq3}
\eea
Trace of ADM propagation:
\bea
   \fl - 3 {1 \over {\cal N}}
       \left( {\dot a \over a} \right)^{\displaystyle\cdot}
       - 3 {\dot a^2 \over a^2}
       - 4 \pi G \left( \varrho + 3 {p \over c^2} \right)
       + \Lambda c^2
       + {1 \over {\cal {\cal N}}} \dot \kappa
       + 2 {\dot a \over a} \kappa
       + {c^2 \Delta {\cal N} \over a^2 {\cal N} (1 + 2 \varphi)}
   \nonumber \\
   \fl \qquad
       = {1 \over 3} \kappa^2
       + 8 \pi G \left( \varrho + {p \over c^2} \right)
       \left( \gamma^2 - 1 \right)
       - {c \over a^2 {\cal N} (1 + 2 \varphi)} \left(
       \chi^{i} \kappa_{,i}
       + c {\varphi^{,i} {\cal N}_{,i} \over 1 + 2 \varphi} \right)
       + c^2 \overline{K}^i_j \overline{K}^j_i.
   \label{eq4}
\eea
Tracefree ADM propagation:
\bea
   \fl \left( {1 \over {\cal N}} {\partial \over \partial t}
       + 3 {\dot a \over a}
       - \kappa
       + {c \chi^{k} \over a^2 {\cal N} (1 + 2 \varphi)} \nabla_k \right)
       \Bigg\{ {c \over a^2 {\cal N} (1 + 2 \varphi)}
   \nonumber \\
   \fl \qquad
       \times
       \left[
       {1 \over 2} \left( \chi^i_{\;\;,j} + \chi_j^{\;,i} \right)
       - {1 \over 3} \delta^i_j \chi^k_{\;\;,k}
       - {1 \over 1 + 2 \varphi} \left( \chi^{i} \varphi_{,j}
       + \chi_{j} \varphi^{,i}
       - {2 \over 3} \delta^i_j \chi^{k} \varphi_{,k} \right)
       \right] \Bigg\}
   \nonumber \\
   \fl \qquad
       - {c^2 \over a^2 ( 1 + 2 \varphi)}
       \left[ {1 \over 1 + 2 \varphi}
       \left( \nabla^i \nabla_j - {1 \over 3} \delta^i_j \Delta \right) \varphi
       + {1 \over {\cal N}}
       \left( \nabla^i \nabla_j - {1 \over 3} \delta^i_j \Delta \right) {\cal N} \right]
   \nonumber \\
   \fl \qquad
       =
       8 \pi G \left( \varrho + {p \over c^2} \right)
       \left[ {\gamma^2 v^i v_j \over c^2 (1 + 2 \varphi)}
       - {1 \over 3} \delta^i_j \left( \gamma^2 - 1 \right)
       \right]
       + {c^2 \over a^4 {\cal N}^2 (1 + 2 \varphi)^2}
   \nonumber \\
   \fl \qquad
       \times
       \Bigg[
       {1 \over 2} \left( \chi^{i,k} \chi_{j,k}
       - \chi_{k,j} \chi^{k,i} \right)
       + {1 \over 1 + 2 \varphi} \left(
       \chi^{k,i} \chi_k \varphi_{,j}
       - \chi^{i,k} \chi_j \varphi_{,k}
       + \chi_{k,j} \chi^k \varphi^{,i}
       - \chi_{j,k} \chi^i \varphi^{,k} \right)
   \nonumber \\
   \fl \qquad
       + {2 \over (1 + 2 \varphi)^2} \left(
       \chi^{i} \chi_{j} \varphi^{,k} \varphi_{,k}
       - \chi^{k} \chi_{k} \varphi^{,i} \varphi_{,j} \right) \Bigg]
       - {c^2 \over a^2 (1 + 2 \varphi)^2}
   \nonumber \\
   \fl \qquad
       \times
       \Bigg[ {3 \over 1 + 2 \varphi}
       \left( \varphi^{,i} \varphi_{,j}
       - {1 \over 3} \delta^i_j \varphi^{,k} \varphi_{,k} \right)
       + {1 \over {\cal N}} \left(
       \varphi^{,i} {\cal N}_{,j}
       + \varphi_{,j} {\cal N}^{,i}
       - {2 \over 3} \delta^i_j \varphi^{,k} {\cal N}_{,k} \right) \Bigg].
   \label{eq5}
\eea
Covariant energy conservation:
\bea
   \fl
       \left[ {\partial \over \partial t}
       + {1 \over a ( 1 + 2 \varphi )} \left( {\cal N} v^k
       + {c \over a} \chi^k \right) \nabla_k \right] \varrho
       + \left( \varrho + {p \over c^2} \right)
       \Bigg\{
       {\cal N} \left( 3 {\dot a \over a} - \kappa \right)
   \nonumber \\
   \fl \qquad
       +
       {({\cal N} v^k)_{,k} \over a (1 + 2 \varphi)}
       + {{\cal N} v^k \varphi_{,k} \over a (1 + 2 \varphi)^2}
       + {1 \over \gamma}
       \left[ {\partial \over \partial t}
       + {1 \over a ( 1 + 2 \varphi )} \left( {\cal N} v^k
       + {c \over a} \chi^k \right) \nabla_k \right] \gamma \Bigg\}
       = 0.
   \label{eq6}
\eea
Covariant momentum conservation:
\bea
   \fl {1 \over a \gamma}
       \left[ {\partial \over \partial t}
       + {1 \over a ( 1 + 2 \varphi )} \left( {\cal N} v^k
       + {c \over a} \chi^k \right) \nabla_k \right]
       \left( a \gamma v_i \right)
       + v^k \nabla_i \left( {c \chi_k \over a^2 ( 1 + 2
       \varphi)} \right)
   \nonumber \\
   \fl \qquad
       + {c^2 \over a} {\cal N}_{,i}
       - \left( 1 - {1 \over \gamma^2} \right) {c^2 {\cal N}
       \varphi_{,i} \over a (1 + 2 \varphi)}
   \nonumber \\
   \fl \qquad
       + {1 \over \varrho + {p \over c^2}}
       \left\{
       {{\cal N} \over a \gamma^2} p_{,i}
       + {v_i \over c^2}
       \left[ {\partial \over \partial t}
       + {1 \over a ( 1 + 2 \varphi )} \left( {\cal N} v^k
       + {c \over a} \chi^k \right) \nabla_k \right] p \right\}
       = 0.
   \label{eq7}
\eea
\textcolor{blue}{ADM energy conservation:}
\bea
   \fl {1 \over {\cal N}} \left[ \varrho
       + \left( \varrho + {p \over c^2} \right)
       \left( \gamma^2 - 1 \right) \right]^{\displaystyle\cdot}
       + {c \over a^2 {\cal N}} {\chi^i \over 1 + 2 \varphi}
       \left[ \varrho
       + \left( \varrho + {p \over c^2} \right)
       \left( \gamma^2 - 1 \right) \right]_{,i}
   \nonumber \\
   \fl \qquad
       + \left( \varrho + {p \over c^2} \right)
       \left(3 {\dot a \over a} - \kappa \right)
       {1 \over 3} \left( 4 \gamma^2 - 1 \right)
       + \left( {\varrho + {p / c^2}
       \over a ( 1 + 2 \varphi )} \gamma^2 v^i \right)_{,i}
   \nonumber \\
   \fl \qquad
       +\left( {3 \varphi_{,i} \over 1 + 2 \varphi}
       + 2 {{\cal N}_{,i} \over {\cal N}} \right)
       {\varrho + {p / c^2} \over a ( 1 + 2 \varphi )}
       \gamma^2 v^i
   \nonumber \\
   \fl \qquad
       + {\gamma^2 \left( \varrho + {p / c^2} \right)
       \over c a^2 {\cal N} ( 1 + 2 \varphi)^2}
       \left[ \chi^{i,j} v_{i} v_{j}
       - {1 \over 3} \chi^j_{\;\; ,j} v^i v_{i}
       - {2 \over 1 + 2 \varphi}
       \left( v^i v^j \chi_i \varphi_{,j}
       - {1 \over 3} v^i v_{i} \chi^j \varphi_{,j} \right)
       \right]
       = 0.
   \label{eq6-ADM}
\eea
\textcolor{blue}{ADM momentum conservation:}
\bea
   \fl \left( {1 \over {\cal N}}
       {\partial \over \partial t}
       + 3 {\dot a \over a} -\kappa \right)
       \left[ a \left( \varrho + {p \over c^2} \right) \gamma^2
       v_{i} \right]
       + {c \over a^2 {\cal N}} {\chi^j \over 1 + 2 \varphi}
       \left[ a \left( \varrho + {p \over c^2} \right)
       \gamma^2 v_{i} \right]_{|j}
   \nonumber \\
   \fl \qquad
       + p_{,i}
       + c^2 \left( \varrho + {p \over c^2} \right)
       {{\cal N}_{,i} \over {\cal N}}
       + \left( {\varrho + {p / c^2} \over 1 + 2 \varphi}
       \gamma^2 v^j v_{i} \right)_{|j}
       + {c \over a {\cal N}} \left( {\chi^j \over 1 + 2 \varphi} \right)_{|i}
       \left( \varrho + {p \over c^2} \right)
       \gamma^2 v_{j}
   \nonumber \\
   \fl \qquad
       + \left[ {1 \over 1 + 2 \varphi} (3 v_{i} \varphi_{,j}
       - v_{j} \varphi_{,i} )
       + {1 \over {\cal N}} ( v_{i} {\cal N}_{,j}
       + v_{j} {\cal N}_{,i} ) \right]
       {\varrho + {p / c^2} \over 1 + 2 \varphi} \gamma^2 v^j
       = 0.
   \label{eq7-ADM}
\eea
With
\bea
   \fl \overline{K}^i_j \overline{K}^j_i
       = {1 \over a^4 {\cal N}^2 (1 + 2 \varphi)^2}
       \Bigg\{
       {1 \over 2} \chi^{i,j} \left( \chi_{i,j} + \chi_{j,i} \right)
       - {1 \over 3} \chi^i_{\;\;,i} \chi^j_{\;\;,j}
       - {4 \over 1 + 2 \varphi} \bigg[
       {1 \over 2} \chi^i \varphi^{,j} \left(
       \chi_{i,j} + \chi_{j,i} \right)
   \nonumber \\
   \fl \qquad
       - {1 \over 3} \chi^i_{\;\;,i} \chi^j \varphi_{,j} \bigg]
       + {2 \over (1 + 2 \varphi)^2} \left(
       \chi^{i} \chi_{i} \varphi^{,j} \varphi_{,j}
       + {1 \over 3} \chi^i \chi^j \varphi_{,i} \varphi_{,j} \right) \Bigg\}.
   \label{K-bar-eq}
\eea

%%%%%%%%%%%%%%%%%%%%%%%%%%%%%%%%%%%%%%%%%%%%%%%%%%%%%%%%%%%%%%%
%
% Bibliography
%
%%%%%%%%%%%%%%%%%%%%%%%%%%%%%%%%%%%%%%%%%%%%%%%%%%%%%%%%%%%%%%%

%%%%%%%%%%%%%%%%%%%%%%%%%%%%%%%%%%%%%%%%%%%%%%%%%%%%%%%%%%%%%%%

%%%%%%%%%%%%%%%%%%%%%%%%%%%%%%%%%%%%%%%%%%%%%%%%%%%%%%%%%%%%%%%
\end{document}